\documentstyle[aps,preprint,tighten,floats]{revtex}
\newcommand{\beq}    {\begin{equation}}
\newcommand{\eeq}    {\end{equation}}
\newcommand{\beqarr} {\begin{eqnarray}}
\newcommand{\eeqarr} {\end{eqnarray}}
\newcommand{\barr}   {\begin{array}}
\newcommand{\earr}   {\end{array}}
\newcommand{\tb}     {{\rm tan} \beta}

\newcommand{\lsim}{\mathrel{\mathop{\kern 0pt \rlap
  {\raise.2ex\hbox{$<$}}}
  \lower.9ex\hbox{\kern-.190em $\sim$}}}
\newcommand{\gsim}{\mathrel{\mathop{\kern 0pt \rlap
  {\raise.2ex\hbox{$>$}}}
  \lower.9ex\hbox{\kern-.190em $\sim$}}}

\begin{document}
\preprint{
\begin{tabular}{r}
CERN--TH 96--103
\\
DFTT 17/96
\\
GEF--Th--5/96
\end{tabular}
}

\title{Relic neutralinos in a best-fitted MSSM}

\author{\bf
A. Bottino$^{\mbox{a,b}}$
\footnote{
E--mail: bottino@to.infn.it,
mignola@vxcern.cern.ch, olechowski@to.infn.it,
scopel@ge.infn.it},
G. Mignola$^{\mbox{c,b}}$,
M. Olechowski$^{\mbox{a,b}}$
\footnote{On leave of absence from the Institute of Theoretical Physics, Warsaw
University, Poland.}
and
S. Scopel$^{\mbox{d,e}}$}
\vspace{1 mm}
\address{
\begin{tabular}{c}
$^{\mbox{a}}$
Dipartimento di Fisica Teorica, Universit\`a di Torino,
Via P. Giuria 1, 10125 Turin, Italy
\\
$^{\mbox{b}}$
INFN, Sezione di Torino,
Via P. Giuria 1, 10125 Turin, Italy
\\
$^{\mbox{c}}$
Theoretical Physics Division, CERN, CH--1211 Geneva 23, Switzerland
\\
$^{\mbox{d}}$
Dipartimento di Fisica, Universit\`a di Genova,
Via Dodecaneso 33, 16146 Genoa, Italy
\\
$^{\mbox{e}}$
INFN, Sezione di Genova,
Via Dodecaneso 33, 16146 Genoa, Italy
\end{tabular}
}
\maketitle

\vspace{2truecm}

\begin{abstract}
We examine the properties of relic neutralinos in the regions of 
the parameter space  
selected by recent fits to all the electroweak observables 
within the Minimal Supersymmetric Standard Model. We discuss the relic 
neutralino cosmic abundance and the direct 
detection rates for these most likely supersymmetric configurations. We 
employ the relevant experimental bounds to
constrain the set of the best-fit configurations and discuss the discovery
potential of the direct search for relic neutralinos. 
\end{abstract}  
\vspace{3truecm}
\noindent
CERN--TH 96--103 \hfill \break
April 1996
\newpage

\section{Introduction}

  The Minimal Supersymmetric extension of the Standard Model (MSSM) has been
analysed by many authors in a variety of different theoretical schemes.
These various scenarios range from the most model-independent one, 
whose parameters at the $M_Z$ scale are constrained only by
experimental bounds, with no assumptions about properties at higher scales, 
to more sophisticated ones,  where the quantities at $M_Z$ are derived from a 
few parameters at the Grand Unification (GU)
scale ($M_{GUT} = O(10^{16})$ GeV),  with the further requirement that the 
ElectroWeak Symmetry Breaking (EWSB) occurs radiatively.

The specific hypotheses that are assumed at $M_{GUT}$ are very crucial for
the phenomenological properties  of the supersymmetric particles at the
$M_Z$  scale.  One often assumes that the soft-breaking mass parameters (scalar
masses, gaugino masses and trilinear couplings) unify at $M_{GUT}$ as well as
the gauge couplings and the $b$--$\tau$ Yukawa couplings. It is worth remarking
that relaxing some of these unification requirements may modify substantially
the supersymmetric phenomenology \cite{OP,others}. The implications of a
relaxation of the soft scalar mass unification at $M_{GUT}$ on the properties 
of relic neutralinos (relic abundance and detection rates) are discussed in 
Refs.\cite{BBEFMS,BBEFMS2}. 

Unfortunately, at present we do not have enough information from 
experimental data on which would be the most reliable supersymmetric 
scheme. In view of this situation, particularly  attractive is 
the approach developed recently, where the parameters of the MSSM at $M_Z$ 
are considered as independent variables in a global fit to the 
electroweak data without any further theoretical assumption 
\cite{inc,alta,GS,CP,KSW,CP2}. 
The fit to all the available experimental data singles out some 
specific regions of the supersymmetric parameter space as the most likely 
ones. Furthermore, this
analysis, which is performed at the $M_Z$ scale,  may provide useful 
insight into the most plausible theoretical scenario at  the $M_{GUT}$ 
scale.

In the present note we examine which phenomenological consequences  
this global fit  entails, within the MSSM, for relic neutralinos. For the most 
likely  domains 
of the parameter space we discuss the features of the relevant neutralino 
relic abundance and of the event rates for the direct search.

\section{$\chi^2$ analysis within the MSSM}

In the global fits of the electroweak data performed in the
framework of the MSSM a key role is played by 
the ratio 
$R_b \equiv \Gamma(Z \rightarrow \bar b b)/ \Gamma(Z \rightarrow \mbox{hadrons}
)$ 
whose experimental value shows a discrepancy (at the level of 3.2
standard deviations) with respect to the value predicted by the Standard Model 
(SM). In fact,  the experimental result 
$R_b^{exp} = 0.2211 \pm 0.0016$ \cite{exp} has to be confronted to the SM 
prediction $R_b^{SM} = 0.2160$ for $m_t = 170$ GeV 
 (or $R_b^{SM} = 0.2158$ for $m_t = 180$ GeV). 
 The MSSM may resolve this disagreement (at least partially), since the 
contributions of some supersymmetric loop diagrams add positively to the 
SM prediction \cite{previous,alta,GS,CP,KSW,CP2}. More specifically, 
substantial supersymmetric contributions occur in two regions of the 
supersymmetric parameter space: (i) for small $\tb$, when both the
chargino and one top squark are light,  and (ii) for large $\tb$, when the
CP-odd Higgs boson $A$ is light and/or both the chargino and one top 
squark are light (also light neutralino and  bottom squark would help in
increasing $R_b$). As usual $\tb$ is defined as $\tb = v_2/v_1$, where $v_1$ and 
$v_2$ are the vacuum expectation values of the Higgs doublets $H_1$ and 
$H_2$, which give masses to the down-type and up-type quarks, respectively.

Thus, including $R_b$ in the overall
set of the EW data makes the $\chi^2$ fit within the MSSM superior to the one
within the SM. At the same time, the experimental range for $R_b$, combined
with other important inputs ($b \rightarrow s \gamma$, bound on $m_h$, etc.) is 
very instrumental in placing stringent bounds on the supersymmetric parameter 
space. Due to the properties recalled above, it is natural that the best fit 
of  the electroweak data within the MSSM automatically selects either 
(i) very small values of $\tb$ (i.e. close to the quasi-IR fixed point for a 
given value of $m_t$) or (ii) very large values of $\tb$ (i.e. of order 
$m_t/{m_b}$) with appropriate  ranges for the masses of 
 $A$ boson,  chargino,   top squark,  neutralino and  bottom
squark \cite{GS,CP,KSW,CP2}. 

It is worth recalling that the fit within the MSSM also has the advantage of 
providing a value for the strong coupling constant $\alpha_s(M_Z)$, which is 
lower than the one found with a SM fit, and  in agreement with the 
measurements of $\alpha_s$ at $q^2 \ll M_Z^2$ \cite{EL,RS,S}. 
Actually, this  is 
expected, since lower values of $\alpha_s(M_Z)$ and additional contributions 
to the $Z$ hadronic width are  related properties \cite{GS,CP,KSW,GS1}. 

A few words of caution are in order here. First, also the experimental value 
of the ratio 
$R_c \equiv \Gamma(Z \rightarrow \bar c c)/ \Gamma(Z \rightarrow \mbox{hadrons}
)$: $R_c^{exp} = 0.1598\, \pm \,0.0070$ \cite{exp} 
shows a disagreement (at the level of 1.7 standard deviations) with the SM
prediction: $R_c^{SM} = 0.172$.  Most disturbingly,  this disagreement in $R_c$ 
cannot be settled within the MSSM. 
These problems affecting $R_b$ and $R_c$ could be due to systematic errors in
the experimental measurements, and  
might therefore disappear after the analysis of new data. 
However, it has to be pointed out that, 
even if $R_c$ is constrained to the SM value, still the measured value of 
$R_b$ shows a disagreement of  $2.6$ standard deviations with the 
SM value \cite{exp}.
In the present note we focus our attention  on the implications that the 
$R_b$ problem would have for relic neutralinos, if the measured effect were 
real and could be accounted for, at least partially, by supersymmetric 
effects within the MSSM.

As was mentioned above, an important ingredient for a good $\chi^2$ at small 
$\tb$ is provided by a positive supersymmetric contribution to $R_b$, when the
chargino is light. This feature is somewhat weakened \cite{ELN}
by the new LEP 1.5 lower bound on the chargino mass 
$m_{\chi^{\pm}} \geq 65~ \mbox{GeV}$ 
(if $m_{\chi^{\pm}} - m_{\chi} \gsim 10$ GeV, $m_{\chi}$ denotes the neutralino
mass) \cite{LEPday}.
However, this new limit does not modify appreciably the conclusions 
of the $\chi^2$
analysis of Ref.\cite{CP}, which we are going to use in the present note, 
since the lowest values of $\chi^2$ at small $\tb$ occur at 
larger values of the chargino mass, $m_{\chi^{\pm}} \simeq$ (75--90) GeV 
\cite{CP2}. 

As far as large values of $\tb$ are concerned, one should recall that 
significant constraints may potentially  follow from the 
$Z \rightarrow b \bar b A$ process \cite{DZZ}. 
This argument was used in Ref.\cite{WK} to conclude that large values of 
$\tb$ are disallowed by present experimental data. However, much larger
statistics and/or more refined analyses of the LEP experimental data on four 
$b$-jets are still required, before solid conclusions on this point may be 
drawn (see also Ref.\cite{CP2}). Here  we consider the large-$\tb$ 
sector of the parameter space as a region deserving much attention, 
because of the indications coming from the fit within the MSSM and 
 in view of the very interesting perspectives that it offers for 
further investigation both at accelerators and with other experimental means 
(direct and indirect searches for relic neutralinos). 

In the analysis presented in this note we take as representative values for case
(i), i.e. small  $\tb$, and case (ii), i.e. large $\tb$, the values: 
$\tb = 1.4$ and $\tb = 50$, respectively. 

Now we briefly mention the main features of the best fit of Ref.\cite{CP}, 
which are relevant to our subsequent discussion on the neutralino 
phenomenology \cite{private}. 
Here all $Z$ partial-widths and  asymmetries at the $Z$ pole are 
calculated and fitted using as 
free parameters the Higgs-mixing parameter $\mu$, the ratio $M_2/\mu$ 
($M_2$ is the SU(2) gaugino mass), the mass of the 
CP-odd Higgs boson  $m_A$, the three soft scalar masses for the third family, 
the soft top trilinear coupling $A_t$ and the strong gauge coupling $\alpha_s$. 
The usual relation between U(1) and SU(2) gaugino masses 
$M_1 =  \frac{5}{3}~ {\rm tan}^2 \theta_W M_2$ is assumed. For details 
concerning the best-fit procedure we refer to \cite{CP,CP2}. 
Various values for the top quark mass were considered in the fit of 
Ref.\cite{CP}. In the present note we set it to $m_t = 170$ GeV. 

For small $\tb$ ($\tb = 1.4$) the best fit ($\chi^2 \simeq 
12$--$13$) does not occur for the smallest possible value of the chargino mass, but
rather for the range $m_{\chi^{\pm}} \simeq$ (75--90) GeV. In fact, for 
such chargino masses the ratio $R_b$ obtains its largest value, 
$R_b \simeq 0.2180$, from the properties of the couplings occurring in the
supersymmetric loop corrections to $R_b$, for one sign of $\mu$, as explained
in Ref. \cite{CP2}. In the configurations 
providing the best fit ($\chi^2 \simeq 12$--$13$), the lighter stop 
is very light indeed: $M_{\tilde t_2} \simeq 50$ GeV, 
the neutralino mass turns out to be constrained in the range 
$m_{\chi} \simeq$ (25--40) GeV, and the mass of the Higgs boson $A$ is larger than
about 200 GeV. If we consider all the configurations that satisfy 
the condition $\chi^2 \leq 16$, then the chargino and neutralino masses are
in the  ranges: 70 GeV $\lsim m_{\chi^{\pm}} \lsim$ 200 GeV and 
25 GeV $\lsim m_{\chi} \lsim$ 100 GeV, respectively. The neutralino may have 
quite different compositions, from very pure gaugino-like to the other extreme of
very pure higgsino-like. In the next section we will discuss the neutralino
relic density. However, we anticipate here which are the modifications to the
features of the best-fit previously discussed, when this best-fit analysis 
is implemented with the
following constraints: (a) the neutralino is the Lightest Supersymmetric
Particle (LSP), (b) the cosmological bound $\Omega h^2 \leq 1$ is satisfied. 
Condition (a) is imposed here by requiring the sfermion masses  to be larger 
than the neutralino mass $m_{\chi}$ by at least $10 \%$, 
i.e. $m_{\tilde f} \geq 1.1\,\, m_{\chi}$. This requirement 
prevents some propagators in the elastic cross sections from becoming
singular and also avoids possible complications due to co-annihilation 
\cite{griest}. 
It turns out that condition (b) is  very effective, particularly in disallowing 
many configurations, since, at small $\tb$, $\Omega_{\chi} h^2$ is usually very 
large. Thus, adding the two conditions (a) and (b) somewhat worsens the 
best-fit quality: the minimal 
$\chi^2$ becomes 13.5 at $m_{\chi^{\pm}} \simeq 90$ GeV and $m_A \simeq 100$ 
GeV. In that region of the parameter space $m_{\chi} \simeq 40~\mbox{GeV}$.
This last value is expected, because the cosmological bound is satisfied due to the 
very pronounced enhancement of the neutralino pair
annihilation cross-section at the $Z$-pole ($m_{\chi} \simeq m_Z/2$). 
The region of the parameter space satisfying the condition $\chi^2 \leq 16$ has
chargino and neutralino masses constrained in the ranges 
65 GeV $\lsim m_{\chi^{\pm}} \lsim \mbox{110 GeV}$ and 
25 GeV $\lsim m_{\chi} \lsim $ 80 GeV. 

Let us now turn to the large-$\tb$ case ($\tb = 50$). The best fit 
($\chi^2 \simeq 12$--$13$) occurs for the smallest possible values of the 
 chargino mass and of the $A$-boson mass: $m_{\chi^{\pm}} \simeq 65~\mbox{GeV}$, 
$m_A \simeq 55$ GeV and for a light neutralino, $m_{\chi} \simeq 50$ GeV.
This allows large values for $R_b$, up to $R_b = 0.2181$. Let us stress that
the relevant chargino and neutralino compositions are mainly higgsino-like, as
implied by the mass relation $m_{\chi} \simeq m_{\chi^{\pm}}$. 
If we consider all the configurations that satisfy the condition 
$\chi^2 \leq 16$, then the masses of chargino and neutralino are in
the ranges: $\mbox{65 GeV} \lsim m_{\chi^{\pm}} \lsim \mbox{250 GeV}$,  
$\mbox{50 GeV} \lsim m_{\chi} \lsim \mbox{190 GeV}$. The cosmological condition 
$\Omega h^2 \leq 1$
is not effective in constraining the parameter space here, as is expected 
at large values of $\tb$.

\section{Relic neutralinos}

We now discuss the properties of the relic neutralinos, should the 
supersymmetric parameter space be the one singled out by the previously 
described $\chi^2$ fit. In particular we discuss the neutralino relic
abundance and  detection  rates for the direct search, where 
a relic neutralino may be detected by the energy released  by its elastic 
scattering off a nucleus of an appropriate set-up 
\cite{dirnoi,dirth,direx}.

Some words of caution are in order here, concerning the evaluation of the
direct detection rates. It is well known that all the 
calculations for detection rates, in addition to the uncertainties due to 
the supersymmetric scheme,  are also affected by many uncertainties 
in some astrophysical and cosmological properties. This is the case for the 
neutralino 
velocity distribution in the halo (assumed here to be Maxwellian
 with a r.m.s. velocity of $270 \pm 24~ {\rm km \cdot s^{-1}}$
\cite{ffg}), 
for its escape velocity ($v_{esc} = 650 \pm 200~{\rm km \cdot s^{-1}}$
\cite{lt}) and 
for the velocity of the Sun around the galactic centre 
($v_{\odot} = 232 \pm 20~ {\rm km \cdot s^{-1}}$\cite{ffg}).  All these 
velocities 
are affected by large error bars, which in turn introduce significant 
uncertainties in the detection rates, especially for light neutralinos. 
  Furthermore, an even more serious 
uncertainty  concerns the value of the local (solar neighbourhood) dark 
matter density $\rho_l$. A recent determination of $\rho_l$, based on 
a flattened dark matter distribution and microlensing data, gives the range 
$\rho_l = 0.51_{-0.17}^{+0.21}~{\rm GeV \cdot cm^{-3}}$  \cite{turner}. 
This introduces a central value significantly larger than  
previous determinations, for instance the one of Ref.\cite{flores}: 
$\rho_l = 0.3 \pm 0.1~{\rm GeV \cdot cm^{-3}}$ , and then entails 
larger detection rates. Notice that the error bars affecting $\rho_l$ are 
 quite large.

Once a specific value for the local density of the {\it total} dark 
matter $\rho_l$ is picked up, we are still confronted with the 
problem of assigning a value to the {\it neutralino} local density 
$\rho_{\chi}$. Here, to determine the value of $\rho_\chi$ to be
used in the detection rates, we adopt  the following rescaling recipe 
\cite{gaisser}: for each point of the parameter
space, we take into account the relevant value of the cosmological neutralino
relic density. When $\Omega_\chi h^2$ is larger than a minimal
$(\Omega h^2)_{min}$, compatible with observational data and with large-scale 
structure calculations, we simply put $\rho_\chi=\rho_l$.
When $\Omega_\chi h^2$ turns out  to be less than $(\Omega h^2)_{min}$, 
and then the neutralino may only provide a fractional contribution
${\Omega_\chi h^2 / (\Omega h^2)_{min}} \equiv  \xi$
 to $\Omega h^2$, we take $\rho_\chi = \rho_l \xi$.
The value to be assigned to $(\Omega h^2)_{min}$ is
somewhat arbitrary, in the range 
$0.03 \lsim (\Omega h^2)_{min} \lsim 0.2$. 

In the present paper, as far as the values of the aforementioned astrophysical 
and cosmological parameters are concerned, we use the two sets of values
reported in Table 1. Set I corresponds essentially to the central values for the
various parameters, whereas set II corresponds to those values of the
parameters, which, within the 
relevant allowed ranges,  provide the lowest detection rates 
(once the supersymmetric variables are fixed). Set I is employed 
to discuss the
average predicted rates, whereas set II is used to derive constraints 
on supersymmetric configurations, whenever the theoretical rates exceed 
the present experimental bounds.

\begin{table}
\caption{Values of the astrophysical and cosmological parameters 
relevant to direct detection rates. 
$V_{r.m.s.}$ denotes the root mean square velocity of the neutralino Maxwellian
velocity distribution in the halo; $V_{esc}$ is the neutralino escape velocity
and $V_{\odot}$ is the velocity of the Sun around the galactic centre; 
$\rho_{loc}$ denotes the local dark matter density and $(\Omega h^2)_{min}$ the
minimal value of $\Omega h^2$. The values of set I are the median values of
the various parameters, the values of set II are the extreme values of the
parameters which, within the physical ranges, provide the lowest estimates of 
the direct rates
(once the supersymmetric parameters are fixed).
}
\begin{center}
\begin{tabular}{|c|c|c|}   \hline
 &  Set I &  Set II \\ \hline
$V_{r.m.s}(\rm km \cdot s^{-1}$) & 270 & 245 \\ \hline
$V_{esc}(\rm km \cdot s^{-1}$)   & 650 & 450 \\ \hline
$V_{\odot}(\rm km \cdot s^{-1}$) & 232 & 212 \\ \hline
$\rho_{loc}(\rm GeV \cdot cm^{-3}$) & 0.5 & 0.2 \\ \hline
$(\Omega h^2)_{min}$            & 0.1 & 0.2 \\ \hline
\end{tabular}
\end{center}
\end{table}

As a further source of uncertainties affecting the detection rates, we 
finally mention the strength of the Higgs--nucleon coupling, which plays an 
important role in the neutralino--nucleus scattering. 
Here, for the relevant coupling  we adopt the determination reported in Ref.
\cite{BBEFMS}.

Many experiments for direct detection of massive weak-interacting dark matter 
particles are under
way, using various materials \cite{direx}. In the present note we consider
detection rates for germanium detectors \cite{Ge,got,beck,cosme,twin}. 
The most consistent procedure for comparing predicted detection rates with 
experimental data is to consider the  differential rate $d R_{direct}/d E_{ee}$
over the whole range of the electron-equivalent energy $E_{ee}$ (or,
alternatively, of the nuclear recoil energy). 
Here, to simplify the discussion, we present our results only in terms of 
an appropriate integrated rate $R_{direct}$. 
On the basis of the experimental spectra of Ref.\cite{twin} we derive, as 
the most stringent upper bound (at 90\% C.L.) for light neutralinos 
(20 GeV $\lsim m_{\chi} \lsim \mbox{50 GeV}$), the value \cite{amor} 

\beq
\int_{6~keV}^{8~keV} dE_{ee}
~dR_{direct}^{exp}/dE_{ee} < 0.9~ \rm events/(kg \cdot day).
\label{eq:a}
\eeq

\noindent
In principle, for heavier neutralinos ($m_{\chi}$ larger than 50 GeV), which are
expected to produce much flatter energy spectra, a more stringent bound 
(at 90\% C.L., still using the data of Ref.\cite{twin}) is
provided by

\beq
\int_{13~keV}^{14~keV} dE_{ee}
~dR_{direct}^{exp}/dE_{ee} < 0.1~ \rm events/(kg \cdot day).
\label{eq:b}
\eeq

\noindent
However, since the limit of Eq. (\ref{eq:b}) introduces only a
marginal improvement over that of Eq. (\ref{eq:a}),   
for simplicity, in the present paper we only use the
bound of Eq. (\ref{eq:a}), which we denote by  $R_{direct}^{exp}$, 
and compare it with the relevant theoretical
integrated rate $R_{direct} \equiv \int_{6~keV}^{8~keV} dE_{ee}~
dR_{direct}/dE_{ee}$.

\section{Results and conclusions}

Let us now turn to the presentation of our results, based on the supersymmetric
configurations selected by the global fit to the electroweak data of 
Refs.\cite{CP,CP2}. We start from the small-$\tb$ case.  In Fig. 1 we show the 
neutralino relic abundance $\Omega_{\chi} h^2$ versus 
$R_{direct}$, for $\tb = 1.4$. As was mentioned in Sect. 2, most of the 
configurations singled out by the best-fit procedure at this value of $\tb$ 
provide large values of $\Omega_{\chi} h^2$, in excess of the cosmological
bound. The scatter plot shown in Fig. 1 displays only those supersymmetric 
configurations that satisfy the condition $\Omega h^2 \leq 1$. The rate 
$R_{direct}$ reported in this figure was evaluated, using set I of 
Table 1. The vertical line corresponds to the experimental bound defined in 
Eq. (\ref{eq:a}). 

We notice in Fig. 1 that many neutralino configurations provide a value for 
the relic abundance within the optimal range for Cold Dark Matter (CDM), 
$0.1 \lsim \Omega_{CDM} h^2 \lsim 0.3$, which is 
suggested by a number of observational data \cite{cosm}. Many other
supersymmetric compositions  provide an even larger contribution to the 
neutralino relic abundance. As expected on general grounds, 
the predicted direct rates are all below the present experimental bound of 
Eq. (\ref{eq:a}). However, it is interesting to note 
that, by improving the experimental sensitivity of the 
direct experiments by about one order of magnitude
(which appears to be within reach in the near future), one can start the 
investigation of a sizeable number of neutralino configurations (including 
many of cosmological interest).

Let us now turn to the large-$\tb$ case. In Fig. 2 we display our results for
the neutralino relic abundance versus the neutralino mass, 
as a scatter plot over all the configurations of the
best-fit analysis with $\chi^2 \leq 16$, 
for $\tb = 50$. 
In this scatter plot the points turn out to be grouped in some characteristic
branches. This feature is due to the grid used in the $\chi^2$ fit: along each
branch the ratio $M_2/|\mu|$ has a fixed value, and each branch stops when 
$m_{\chi^{\pm}}$ reaches the maximal value considered in the fit: 
$m_{\chi^{\pm}}  = 200$ GeV ($m_{\chi^{\pm}}=250$ GeV for the value  
$M_2/|\mu|=1$). 
The highest branch corresponds to the value $M_2/|\mu| = 0.5$ (gaugino
dominance in the neutralino composition), the intermediate one belongs to 
the value $M_2/|\mu| = 1$ (large 
gaugino--higgsino mixing), and the lowest branch contains configurations with 
$M_2/|\mu| = 3, 10$ (higgsino dominance). The different nature of the neutralino
compositions along the different branches explains the significant differences
in their $\Omega_{\chi} h^2$ values. As expected, only gaugino-like neutralinos
may have values of relic abundance in the desirable range 
$0.1 \lsim \Omega_{CDM} h^2 \lsim 0.3$. Actually, only very few configurations
within the total best-fit sample fall into this range. However, it
is rewarding that a large number of states
provide a relic abundance of some cosmological interest, i.e. in the range 
$\Omega_{\chi} h^2 \gsim 0.025$.

Let us now discuss the event rates for direct detection of relic neutralinos. 
Figure 3 displays two scatter plots of the rate $R_{direct}$ as a function of 
$m_{\chi}$. 
The rate $R_{direct}$ shown in section (a) was evaluated 
using set I in Table 1, the one shown in section (b) was 
evaluated using set II. 
In Fig. 3  configurations of different cosmological relevance are denoted by
different symbols.
The shape of the rate $R_{direct}$ in these plots is reminiscent of the shape 
of the relic abundance displayed in Fig. 2, the reason being that the rescaling 
of the local neutralino density is very effective here. Of the two sets 
of points which branch off at $m_{\chi} \simeq 80$ GeV, the upper branch is 
mainly a combination of the configurations with $M_2/|\mu| = 0.5, 1, 3$, 
whereas the 
lower branch contains configurations with $M_2/|\mu| = 10$. In fact, the 
neutralino--nucleus cross section that enters in the detection rate has the 
effect of somewhat grouping together the configurations belonging to the three 
smallest values of $M_2/|\mu|$ and of further suppressing the rates for the 
purest higgsino states with $M_2/|\mu| = 10$. 
Of the two plots shown in Fig. 3, the one of section (a), which was obtained by
using the values of set I for the astrophysical parameters, is indicative of
what would be expected from direct detection, but suffers from 
uncertainties due to the large error bars affecting the various parameters.
The plot of section (b) is the one which is relevant, when we address the 
question whether the present experimental
bound $R_{direct}^{exp}$ already allows us to set some constraints on our
sample of best-fit supersymmetric configurations. 
As a matter of fact, the scatter plot displayed in Fig. 3b represents a
conservative estimate of the predicted rates. Thus we may conclude that the
neutralino configurations, whose rates are above $R_{direct}^{exp}$ in the plot
of section (b), are excluded by the present experimental bound from
direct detection experiments \cite{future}. The
number of the disallowed configurations  is almost one tenth  of the  total 
number of about 3,000 configurations. 

In Fig. 4 we display, in an  $R_b$ versus $\chi^2$ plot, 
the frontier of the region that contains the whole sample of the best-fit 
configurations of 
Ref.\cite{CP} (this frontier is denoted by a solid line). The dashed curve 
delimits the region where are located those configurations of the original 
sample that are excluded by $R_{direct}^{exp}$ on the
basis of the results shown in Fig. 3b.  However, we stress that the region 
inside the dashed curve is still populated by a large number of other allowed
configurations. 

Some noticeable properties emerge from our results:
\begin{itemize}
\item The configurations excluded by the present upper bound 
      $R_{direct}^{exp}$ are not located in the minimum of $\chi^2$ 
      (see Fig. 4).
      Actually, 
      from details of our analysis which are not displayed in Fig. 4, it turns
      out that most of the excluded configurations are concentrated in the 
      range $15 \lsim \chi^2 \lsim 16$. 
      The configurations that give the minimal values 
      of $\chi^2$ have a very pure higgsino composition, and then provide a 
      small detection rate.        
      Therefore, the constraint
      introduced by $R_{direct}^{exp}$, although instrumental in limiting 
      the parameter space, does not nevertheless modify the good quality
      of the fit significantly.

\item Even a modest improvement in the sensitivity of the
      direct search experiments allows the 
      exploration of a large number of neutralino configurations. 
      This is clear from Fig. 3b, which shows that the 
      $R_{direct}$ vs.  $m_{\chi}$ plot is highly populated around the line 
      denoting the present upper bound $R_{direct}^{exp}$. For example, our 
      calculations predict that, once the experimental sensitivity is improved 
      by an order of magnitude with respect to the present one, more than 
      60\% of the total number of the best-fit configurations can be explored.
      Some of them belong to the region of minimal $\chi^2$. 
      Thus, the discovery potential of the direct search for relic neutralinos 
      is quite remarkable. 

\end{itemize}
\vspace{10 mm}

\vspace{10 mm}
{\bf Acknowledgements}

We thank Stefan Pokorski, Piotr Chankowski and Nicolao Fornengo for 
interesting discussions. This work was
supported in part by the Research Funds of the Ministero dell'Universit\`a 
e della Ricerca Scientifica e Tecnologica. A fellowship of the 
Universit\`a di Torino is gratefully acknowledged by G. Mignola. 
M. Olechowski would like to 
acknowledge the support of the Polish Committee of Scientific Research.

\vfill
\eject

{\bf Figure Captions}

\vspace{10 mm}
{\bf Figure 1} - Neutralino relic abundance $\Omega_{\chi} h^2$ versus 
$R_{direct}$, for $\tb = 1.4$. This scatter plot displays only supersymmetric 
configurations that satisfy the condition $\Omega h^2 \leq 1$. The rate 
$R_{direct}$ reported in this figure was evaluated, using set I of 
Table 1. The vertical line corresponds to the experimental bound given in 
Eq. (\ref{eq:a}). 

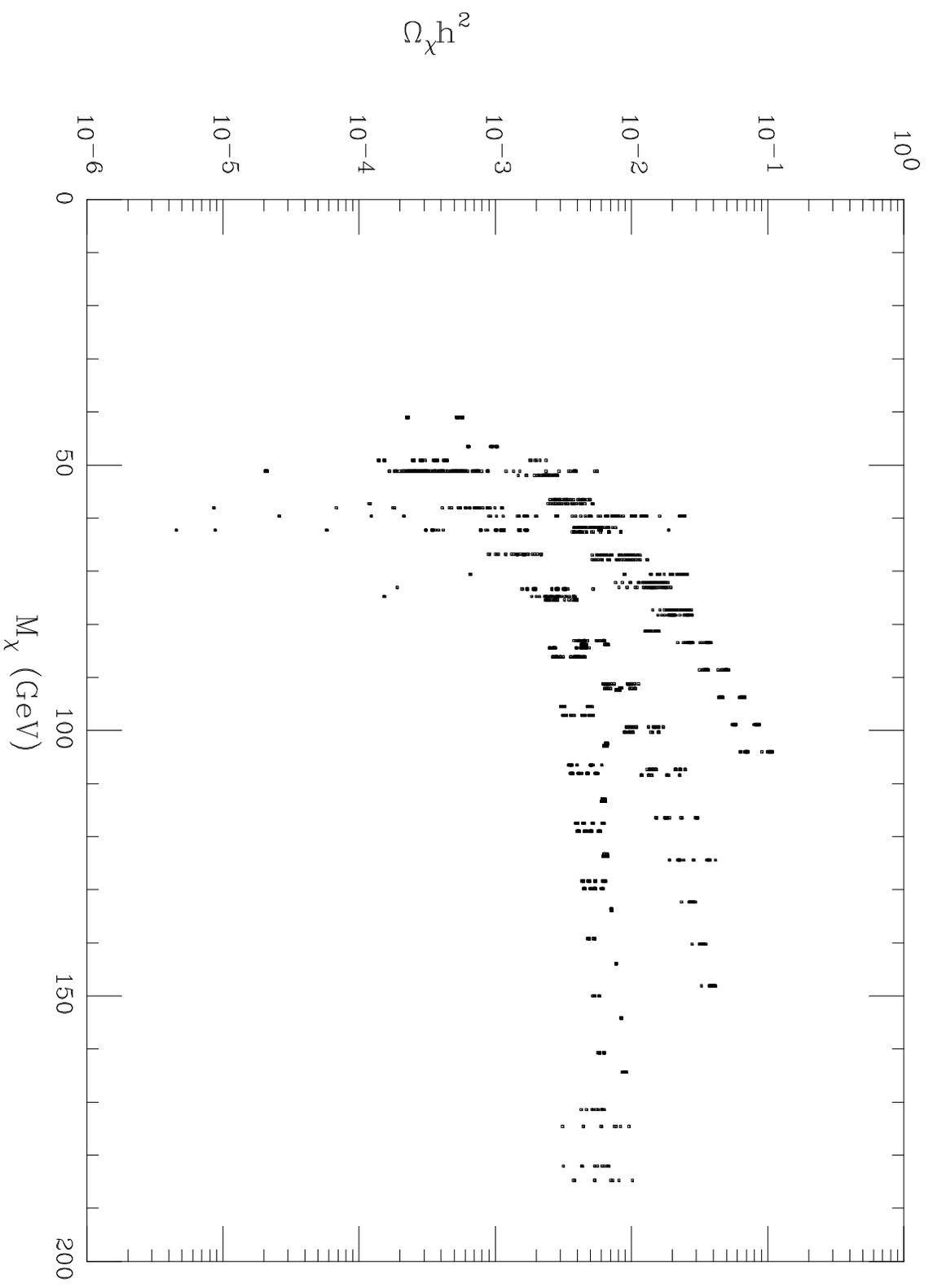
{\bf Figure 2} - Neutralino relic abundance versus the neutralino mass, 
as a scatter plot over all the configurations of the
best-fit analysis, for $\tb = 50$. 

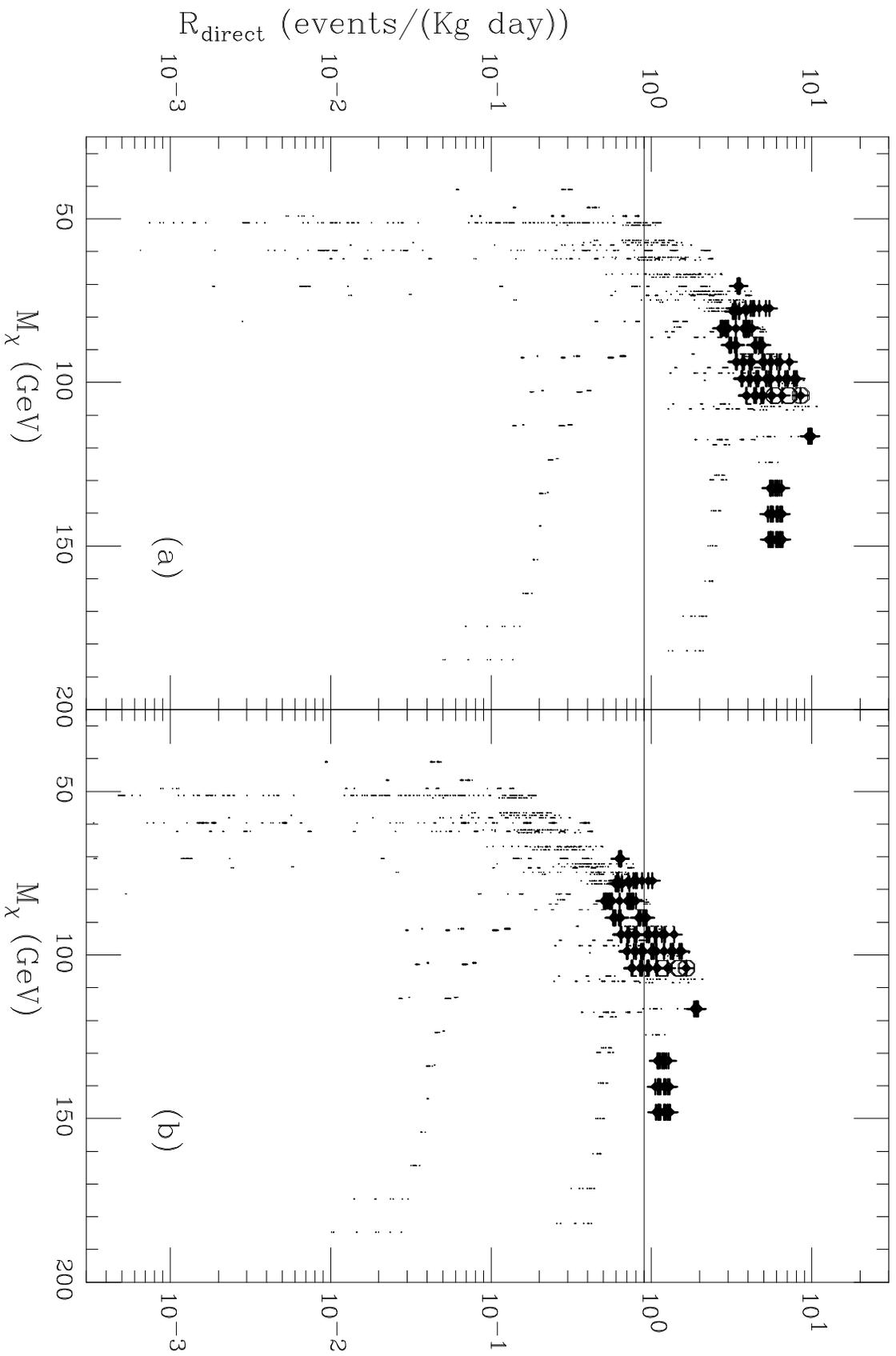
{\bf Figure 3} - Scatter plots of the rate $R_{direct}$ as a function of 
$m_{\chi}$. The rate $R_{direct}$ shown in section (a) has been evaluated 
using set I in Table 1, the rate $R_{direct}$ shown in section (b) 
using set II. 
Configurations of different cosmological relevance are denoted by
different symbols: configurations with a relic abundance 
$\Omega_{\chi} h^2 \geq 0.1$ are denoted by circles, those with a
relic abundance in the range  $0.025 \leq \Omega_{\chi} h^2 < 0.1$ are
denoted by crosses, the other configurations by points. 

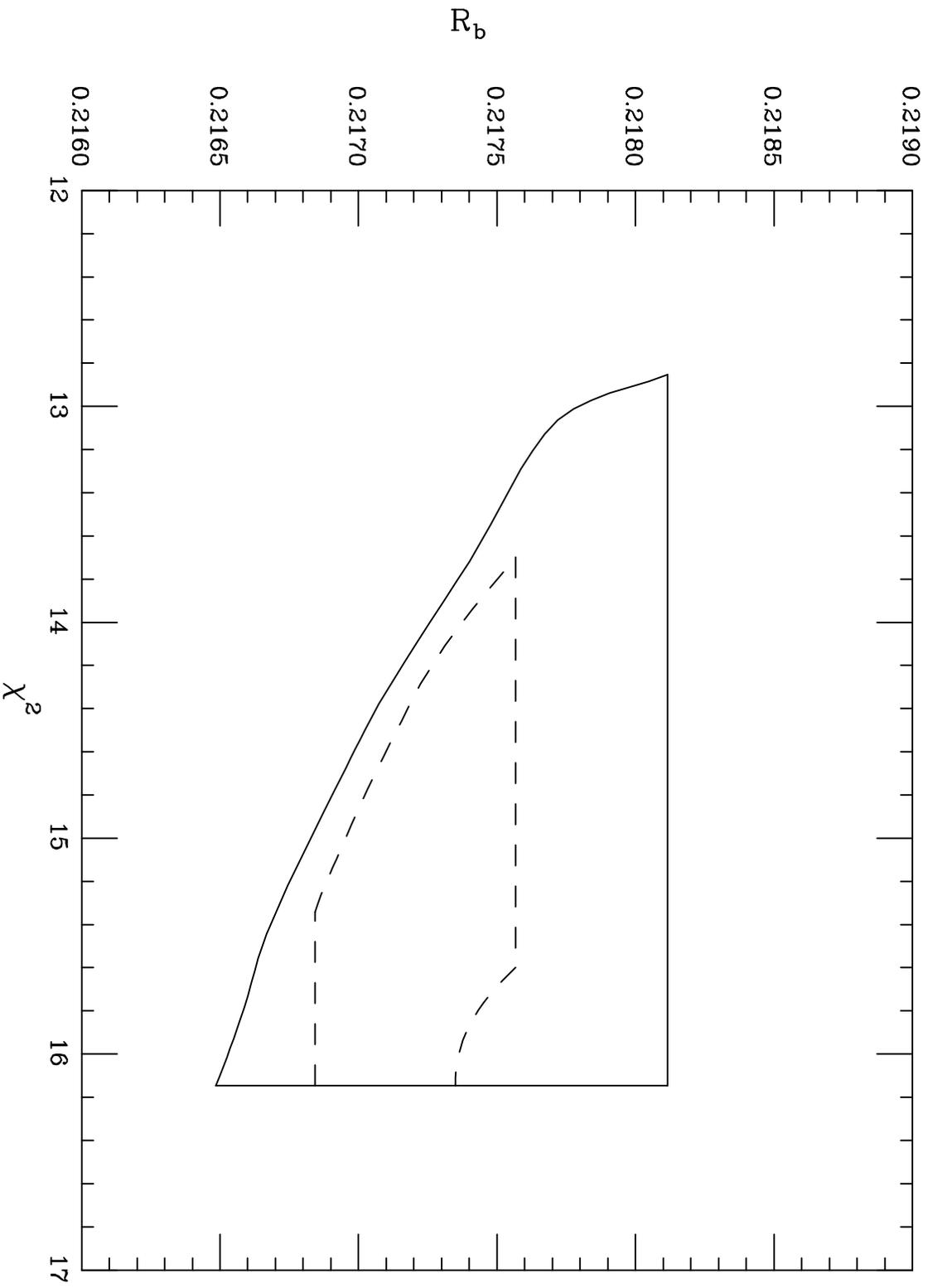
{\bf Figure 4} - Plot of $R_b$ versus $\chi^2$. The solid line delimits 
the region which contains the whole sample of the best-fit 
configurations of 
Ref.\cite{CP}. The dashed curve 
delimits the region where those configurations of the original 
best-fit sample excluded by $R_{direct}^{exp}$ are located.


\begin{thebibliography}{99}



\vspace{10 mm}
\bibitem{OP} M. Olechowski and S. Pokorski,  
  {\it Phys. Lett. } {\bf B344} (1965) 201.


\bibitem{others}
N. Polonsky and A. Pomarol, {\it Phys. Rev. Lett.} {\bf 73} (1994) 2292;
{\it Phys. Rev.} {\bf D51} (1995) 6532; 
D. Matalliotakis and  H. P. Nilles, {\it Nucl. Phys.} {\bf B435}
(1995) 115; 
A. Pomarol and S. Dimopoulos,  {\it Nucl. Phys.} {\bf B453} (1995) 83; 
H. Murayama, Berkeley preprint LBL--36962 (1995), hep-ph/9503392.

\bibitem{BBEFMS} V. Berezinsky, A. Bottino, J. Ellis, N. Fornengo, 
G. Mignola and S. Scopel, CERN preprint CERN-TH 95-206 (to appear in 
{\it Astroparticle Physics}), hep-ph/9508249.

\bibitem{BBEFMS2} V. Berezinsky, A. Bottino, J. Ellis, N. Fornengo, 
G. Mignola and S. Scopel, CERN preprint CERN-TH 96-42, hep-ph/9603342.


\bibitem{inc} J. Ellis, G.L. Fogli and E. Lisi, {\it Phys. Lett.} 
{\bf 286B} (1992) 85, 
{\it Nucl. Phys.} {\bf B393} (1993) 3, {\it Phys. Lett.} {\bf 324B} (1994) 
173 and  {\bf 333B} (1994) 118; P. Langacker and M. Luo, 
{\it Phys. Rev.} {\bf D44} (1991) 817; 
M. Carena and C.E.M. Wagner, {\it Nucl. Phys.} {\bf B452} (1995) 45. 

\bibitem{alta} G. Altarelli, R. Barbieri and F.
Caravaglios, {\it Phys. Lett.} {\bf 314B} (1993) 357. 

\bibitem{GS} D. Garcia and J. Sol\`a, {\it Phys. Lett.} {\bf B354} (1995) 
335.

\bibitem{CP} P. H. Chankowski and S. Pokorski, {\it Phys. Lett.} {\bf 366B}
(1996) 188.

\bibitem{KSW} G.L. Kane, R.G. Stuart and J.D. Wells, 
{\it Phys. Lett.} {\bf B354} (1995) 350.

\bibitem{CP2} P. H. Chankowski and S. Pokorski, preprint IFT-96/6
(hep-ph 9603310).

\bibitem{exp} M.D. Hildreth, talk at the XXXI Rencontres de Moriond, 
March 1996.

\bibitem{previous} A. Djouadi, G. Girardi, C. Verzegnassi, W. Hollik and 
F.M. Renard, {\it Nucl. Phys.} {\bf B349} (1991) 48;
M. Boulware and D. Finnell, {\it Phys. Rev.} {\bf D44} (1991) 2054;
A. Blondel and C. Verzegnassi, {\it Phys. Lett. } {\bf B311} (1993) 346. 


\bibitem{EL} J. Erler and P. Langacker, 
  {\it Phys. Rev.} {\bf D52} (1995) 441.

\bibitem{RS} L. Roszkowski and M. Shifman, {\it Phys. Rev.} {\bf D53} (1996)
404.

\bibitem{S} M. Shifman, {\it Mod. Phys. Lett.} {\bf A10} (1995) 605. 

\bibitem{GS1} D. Garcia and J. Sol\`a, {\it Phys. Lett.} 
{\bf B357} (1995) 349.

\bibitem{ELN} J. Ellis, J.L. Lopez and D.V. Nanopoulos, CERN preprint 
CERN-TH/95-314, hep-ph/9512288.


\bibitem{LEPday} L. Rolandi, H. Dijkstra, D. Strickland and G. Wilson,
representing the ALEPH, DELPHI, L3 AND OPAL collaborations, Joint Seminar 
on the First Results from LEP 1.5, CERN, December 1995; ALEPH Collaboration, 
preprint CERN-PPE/96-10; OPAL Collaboration, preprint CERN-PPE/96-019 and 
CERN-PPE/96-020.

\bibitem{DZZ} A. Djouadi, P. Zerwas and J. Zunft, 
{\it Phys. Lett.} {\bf B259} (1991) 175.
 
\bibitem{WK} J. Wells and G. Kane, {\it Phys. Rev. Lett.} {\bf 76} (1996) 869.

\bibitem{private} We thank  Piotr Chankowski for providing us with the 
outputs of the $\chi^2$ fit of Ref. \cite{CP}, employed in our analysis. 

\bibitem{griest}
K. Griest and D. Seckel, {\it Phys. Rev.} {\bf D 43} (1991) 3191. 

\bibitem{dirnoi} A. Bottino, V. de Alfaro, N. Fornengo, G. Mignola and 
S. Scopel, 
{\it Astroparticle Physics} {\bf 2} (1994) 77; A. Bottino, V. de Alfaro, 
N. Fornengo, G. Mignola, S. Scopel, and \mbox{C. Bacci} et al. 
(BRS Collaboration), {\it Phys. Lett.} {\bf B295} (1992) 330.
See also the review paper by \mbox{G. Jungman}, M. Kamionkowski and K. Griest, 
{\it Phys. Rep.} {\bf 267} (1996) 195, and
references quoted therein. 

\bibitem{dirth} L. Bergstr\"om and P. Gondolo, preprint OUTP-95-38P, 
hep-ph/9510252.

\bibitem{direx} For reviews covering also the experimental aspects, see: 
J. R. Primack, D. Seckel and B. Sadoulet, {\it Annu. Rev. Nucl.
Part. Sci.} {\bf 38} (1988) 751; P. F. Smith and J. D. Lewin, {\it Phys.
Rep.} {\bf 187} (1990) 203; R. Bernabei, {\em Riv. N. Cim.} {\bf 18} (1995) N.5,
and L. Mosca, invited talk at TAUP 95 
(Toledo, September 1995) to appear in {\em Nucl. Phys.} {\bf B} (Proc. Suppl.).



\bibitem{ffg}
F. J. Kerr, D. Lynden-Bell, {\it Mon. Not. R. Astr. Soc.} {\bf 221} (1986)
1023.

\bibitem{lt}
P.J.T. Leonard and S. Tremaine, {\em Ap.J.} {\bf 353} (1990) 486.

\bibitem{turner} E.I. Gates, G. Gyuk and M. Turner, Fermilab preprint 
FERMILAB-Pub-95/090-A, hep-ph/9505039.

\bibitem{flores} R.A. Flores, {\it Phys. Lett.} {\bf B215} (1988) 73.
\bibitem{gaisser} T.K. Gaisser, G. Steigman and S. Tilav, {\it Phys. Rev.} 
{\bf D34} (1986) 2206.

\bibitem{Ge} S. P. Ahlen et al., {\it Phys. Lett.} {\bf B195} (1987) 603; 
D. O. Caldwell et al., {\it Phys. Rev. Lett.} {\bf 61} (1988) 510.

\bibitem{got} D. Reusser et al., {\it Phys. Lett.} {\bf B255} (1991) 143.

\bibitem{beck} M. Beck (Heidelberg--Moscow  Collaboration),
{\it Nucl. Phys.}
{\bf B} (Proc. Suppl.) {\bf35} (1994) 150; M. Beck et al., {\it Phys. Lett.}
 {\bf B336} (1994) 141.

\bibitem{cosme} E. Garcia et al., {\it Nucl. Phys. B} (Proc. Suppl.) 
{\bf 28A} (1992) 286; M. L. Sarsa et al., 
{\it Nucl. Phys.} {\bf B} (Proc. Suppl.) {\bf 35} (1994) 154; 
E. Garcia et al., Proc. ``The Dark Side of the Universe", 
eds. R. Bernabei and C. Tao (World Scientific, Singapore 1994), p. 216.

\bibitem{twin} A.K. Drukier et al., 
{\it Nucl. Phys. B(Proc. Suppl.)} {\bf 28A} (1992) 293; I. R. Sagdev, 
\mbox{A. K. Drukier}, D. J. Welsh, A. A. Klimenko,
S. B. Osetrov, A. A. Smolnikov, 
{\it Nucl. Phys.} {\bf B} (Proc. Suppl.) {\bf 35} (1994) 175. 

\bibitem{amor} We thank Angel
Morales for providing us with the listings of the COSME \cite{cosme} and TWIN 
\cite{twin} germanium spectra.

\bibitem{cosm} For a discussion of this point, see  
Ref.\cite{BBEFMS2} 
and references quoted therein.

\bibitem{future} An analysis, along the general lines of the present paper,
based on experimental bounds on up-going muon fluxes from the centre of the
Earth and from the Sun (indirect neutralino detection) is in preparation. 

\end{thebibliography}
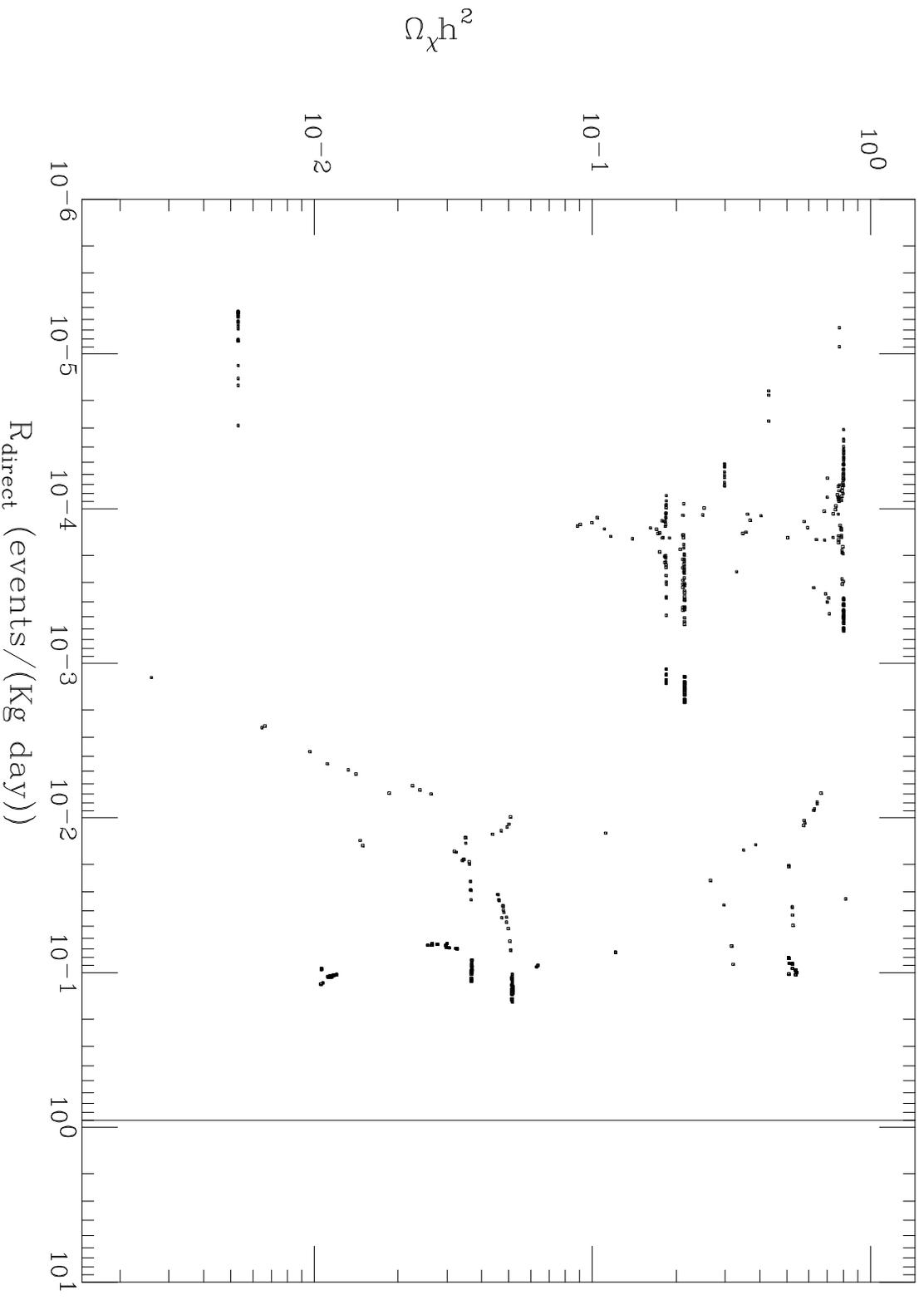
\end{document}